\documentclass[conference]{IEEEtran}
\IEEEoverridecommandlockouts
\usepackage{cite}
\usepackage{amsmath,amssymb,amsfonts}
\usepackage{algorithmic}
\usepackage{graphicx}
\usepackage{textcomp}
\usepackage{xcolor}
\usepackage{graphicx}
\usepackage{color}
\usepackage{placeins}
\usepackage{float}
\usepackage{tabularx,colortbl}
\usepackage{subfigure}
\usepackage{multirow}
\usepackage{booktabs}
\usepackage{makecell}

\def\BibTeX{{\rm B\kern-.05em{\sc i\kern-.025em b}\kern-.08em
    T\kern-.1667em\lower.7ex\hbox{E}\kern-.125emX}}
\begin{document}

\title{Flexible-Region Based Adaptive In-Loop Filter for Video Coding}

%\author{\IEEEauthorblockN{Xuewei Meng}
%\and
%\IEEEauthorblockN{Chuanmin Jia}
%\and
%\IEEEauthorblockN{Shanshe Wang}
%\and
%\IEEEauthorblockN{Siwei Ma}
%\and
%\and

\author{\IEEEauthorblockN{Xuewei Meng\IEEEauthorrefmark{1},
		Chuanmin Jia\IEEEauthorrefmark{1},
		Jing Cui\IEEEauthorrefmark{1},
		Shanshe Wang\IEEEauthorrefmark{1},
        Xiaozhen Zheng\IEEEauthorrefmark{2},
		and
		Siwei Ma\IEEEauthorrefmark{1}}
        \IEEEauthorblockA{\IEEEauthorrefmark{1}Institute of Digital Media, Peking University, Beijing, China\\
        Email: \{xwmeng, cmjia, jingcui106, sswang, swma\}@pku.edu.cn}
        \IEEEauthorblockA{\IEEEauthorrefmark{2}SZ DJI Technologies Co., Ltd. Shenzhen, China\\
        Email: xiaozhen.zheng@dji.com}}
	
\maketitle
\bibliographystyle{IEEEtran}

\begin{abstract}
Adaptive loop filter (ALF) for video coding, which is designed to minimize the mean square error between original and reconstructed samples by using Wiener-based filter, has attracted increasing attention for its significant capability in improving coding efficiency. In the second and third Audio Video Coding Standard, i.e., AVS2 and AVS3, ALF is adopted as one of the in-loop filters. In current design, each frame is divided into 16 regions at most and corresponding filter coefficients are then derived and utilized to reconstruct each region. In this paper, a flexible-region based ALF (FRALF) scheme is proposed to improve the adaptability of existing ALF in AVS3, which introduces multiple region partition templates, such as $2\times4$, $4\times4$, $4\times8$ and $8\times8$. We subsequently propose the filter coefficients merging algorithm to further improve coding efficiency by estimating the distortion level of different partition regions. The proposed FRALF can fully consider the local texture characteristics as well as non-local similarities synthetically. The experimental results show that FRALF outperforms the existing region-based ALF in AVS3 with relatively low complexity increasing. 
\end{abstract}


\section{Introduction}
The state-of-the-art video coding standard of Audio and Video Coding Standard (AVS), AVS2, developed by the IEEE 1857 Working Group under project 1857.4, has been standardized in 2015 \cite{AVSOverview}. AVS2 is the second-generation video coding standard established by the AVS Working Group of China, which can achieve about 50\% bit-rate saving compared to AVS1 \cite{AVS2?}\cite{AVS2framework} and almost the same performance compared to H.265/HEVC (High Efficiency Video Coding) \cite{HEVC}. To further improve the coding efficiency, AVS Working Group is developing the next generation of AVS video coding standard, AVS3, since Jul. 2018. And the reference software, High Performance Model (HPM), was initiated in Dec. 2018 \cite{HPM}.

In the existing representative hybrid coding based video coding standards, i.e., HEVC and AVS2, in-loop filters play an important role in improving coding efficiency by reducing compression artifacts, such as blocking, ringing, and blurring artifacts. During the development of video coding standards, many kinds of in-loop filters are proposed, such as Deblocking Filter (DF) \cite{DF}, Sample Adaptive Offset (SAO) \cite{SAO}, Adaptive Loop Filter~(ALF)~\cite{ALF,NonlinearALF-1,NonlinearALF-2,meng2021optimized,meng2026optimizedadaptiveloopfilter}, Cross-Component Adaptive Loop Filter \cite{OnePass,CC-ALF,CCALF1,CCALF2,r0322,Parallel,R0225}, Bilateral Filter \cite{BF}, Image prior based Filter \cite{zhang2015nonlocal,ma2016nonlocal,SANF,NALF,meng2018optimized,nonlocal2,nonlocal3,nonlocal4}, CNN-based filter \cite{jia2019content} and so on. While in AVS2 and AVS3, considering both performance and complexity, there are three kinds of in-loop filters, DF \cite{DF_AVS}, SAO \cite{SAO_AVS}, and ALF \cite{ALF_AVS}. DF is a non-linear filter with predefined low-pass characteristics to reduce blocking artifacts \cite{ALF} at boundaries of Coding Unit (CU), Prediction Unit (PU) and Transform Unit (TU). Unlike DF, SAO utilizes the original samples of current picture to reduce the Mean Square Error (MSE) between the original and reconstructed samples by adding an offset. Similar with SAO, ALF minimizes the MSE between original and reconstructed image based on wiener filtering principle. ALF can reduce compression artifacts and improve objective coding efficiency significantly. Due to this characteristic, ALF has become an emerging research topic in both industry and academic groups.

The basic idea of ALF is to estimate one or more Wiener-based filters at encoder by minimizing the MSE between original and reconstructed samples and signal the parameters to decoder. From inverse problem perspective, signal-level reconstruction quality could be maximized if each pixel owns its own ALF coefficients. However, it is unrealistic because of the huge overhead caused by transmitting ALF coefficients. Hence, artifacts reduction and overhead should be considered jointly and comprehensively. To solve this problem, there are mainly two kinds of methods, classification-based ALF and region-based ALF. As for classification-based ALF \cite{GALF}, pixels or blocks are classified based on the predefined rules, such as 1-D Laplacian direction and 2D Laplacian activity, and all pixels in the same class share one set of ALF coefficients. Classification-based adaptive loop filter can improve the adaptability of ALF at the expense of higher computational complexity and worse hardware friendliness. Region-based ALF, which is used in AVS2 and AVS3, assumes pixels in the same region have similar characteristics and share the same ALF coefficients. In current design, each frame is divided into 16 regions at most based on fixed rules regardless of its resolution and the 16 regions are merged based on Hilbert-Scan mode. Dividing all frames into 16 regions doesn't fully consider the texture difference within one region. Meanwhile, conducting coefficients merging between neighboring regions by fixed scanning mode doesn't fully consider the non-local similarities between non-adjacent regions.

In this paper, in order to improve the performance of region-based ALF in AVS3 without increasing the complexity in decoder, we propose a flexible-region based adaptive loop filter (FRALF) whose main idea is dividing one frame into multiple regions according to its texture characteristics and conduct coefficients merging between neighboring and non-neighboring regions in each frame, which considering the local texture characteristics as well as non-local similarities synthetically. Our proposed adaptive region partition scheme for ALF achieves up to 0.31\%, 0.36\% and 0.36\% BD-rate saving for All Intra (AI), Random Access (RA) and Low Delay B (LDB) configurations, respectively,  compared to original ALF in HPM-3.3.

The rest of the paper is organized as follows. Section \uppercase\expandafter{\romannumeral2} briefly reviews ALF technique in AVS3. In Section \uppercase\expandafter{\romannumeral3}, the proposed flexible-region based ALF and its implementation details are elaborated. Section \uppercase\expandafter{\romannumeral4} shows experimental results of the proposed method. Finally, conclusions are drawn in section \uppercase\expandafter{\romannumeral5}.

\section{Review of Adaptive Loop Filter in AVS3}
ALF is introduced in AVS2 and AVS3 to minimize the mean square error between original and reconstructed samples by using Wiener-based filter. There are mainly three modules in existing ALF technique of AVS3, including fixed region partition template, hilbert-scan based coefficients merging and Wiener filter derivation, and group based filtering by applying corresponding filter coefficients to neighboring samples. We mainly focus on the first two modules in this paper. The details are given below.

\subsection{Region Partition Strategy}
\begin{figure}[ht]
	\begin{center}
		\noindent
		\subfigure[$416 \times 240$]{
			\includegraphics[width=3.0in]{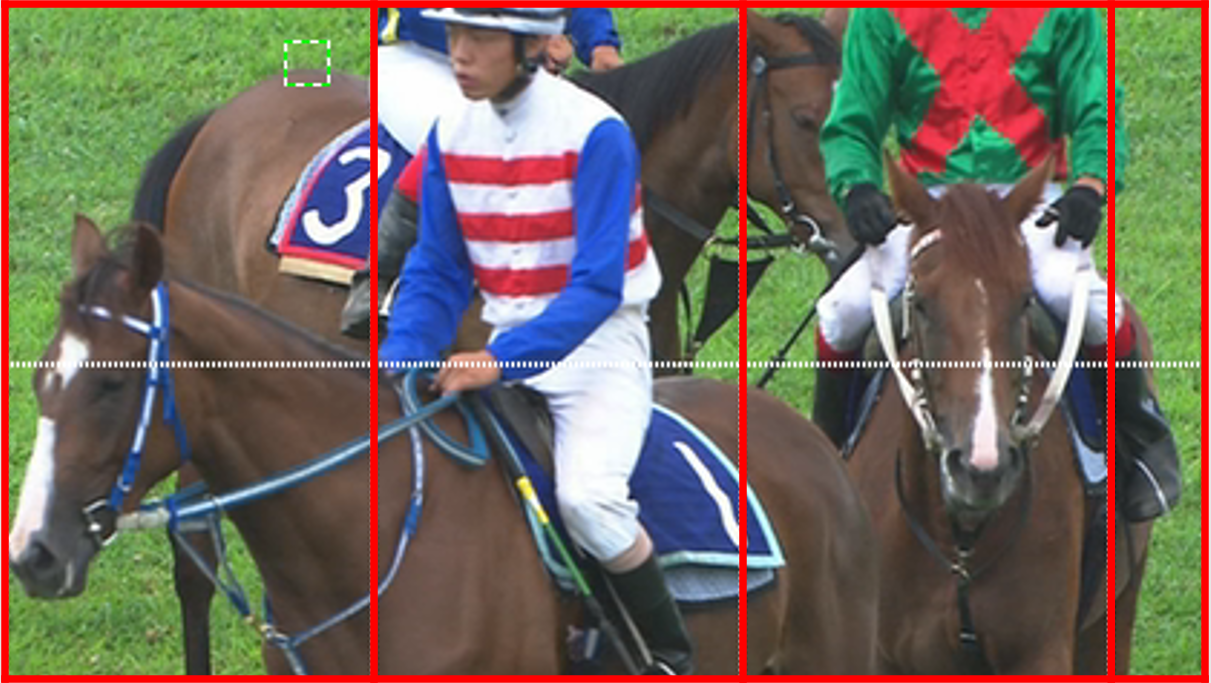}
		}
		\subfigure[$1280 \times 720$]{
			\includegraphics[width=3.0in]{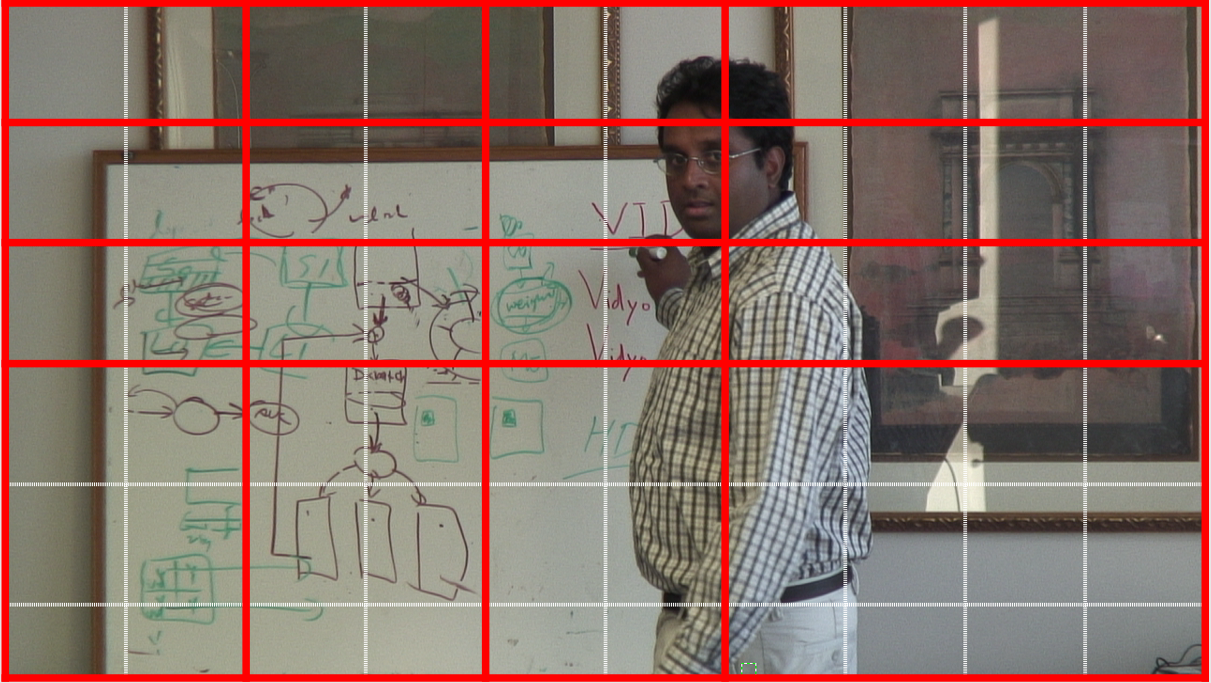}
		}
		\caption{Region partition examples in ALF (AVS3)}\label{region}
	\end{center}
\end{figure}

As shown in Fig.~\ref{region}, each frame is divided into 16 regions by {H} and {W} folds ({H = 4. W = 4}) at most. Each boundary of the region is aligned with the Coding Tree Unit (CTU) boundaries, which means that one CTU can not be divided into two or more different regions. In AVS3, the size of CTU is $128 \times 128$. CTU boundaries are marked by white lines and region boundaries are colored by red in Fig.~\ref{region}. For videos with resolution $416 \times 240$, as shown in Fig.~\ref{region}(a), there are 8 CTUs and 4 regions in each frame according to the existing ALF partition template. As for $1280 \times 720$ videos, as shown in Fig.~\ref{region}(b), each frame is divided into 16 regions. The smallest region owns 2 CTUs, while the largest one has 12 CTUs. It can be seen that the current region partition strategy tends to generate non-uniformly partitioned regions, and in some cases the area of each region is so large that the texture distribution in it is irregular. Such unbalanced region partition mechanism may lead to low filtering performance due to uneven number of samples in each region.

%\begin{equation}
%CTUs\_col = CTU\_NUM_x/W,
%\end{equation}
%
%\begin{equation}
%CTUs\_row = CTU\_NUM_y/H,
%\end{equation}
%
%\begin{equation}
%Region\_pos(i, j) = (CTU\_col \times i \times CTU\_width, CTU\_row \times j \times CTU\_height),
%\end{equation}

\begin{figure}[ht]
	\begin{center}
		\noindent
		\includegraphics[width=2.5in]{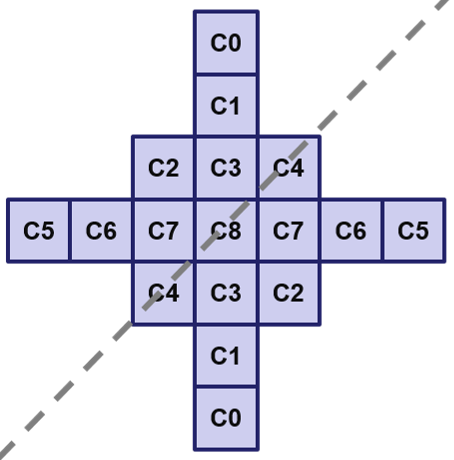}
		\caption{Filter shape used in ALF (AVS3)}\label{filter shape}
	\end{center}
\end{figure}

\begin{figure}[ht]
	\begin{center}
		\noindent
		\includegraphics[width=2.5in]{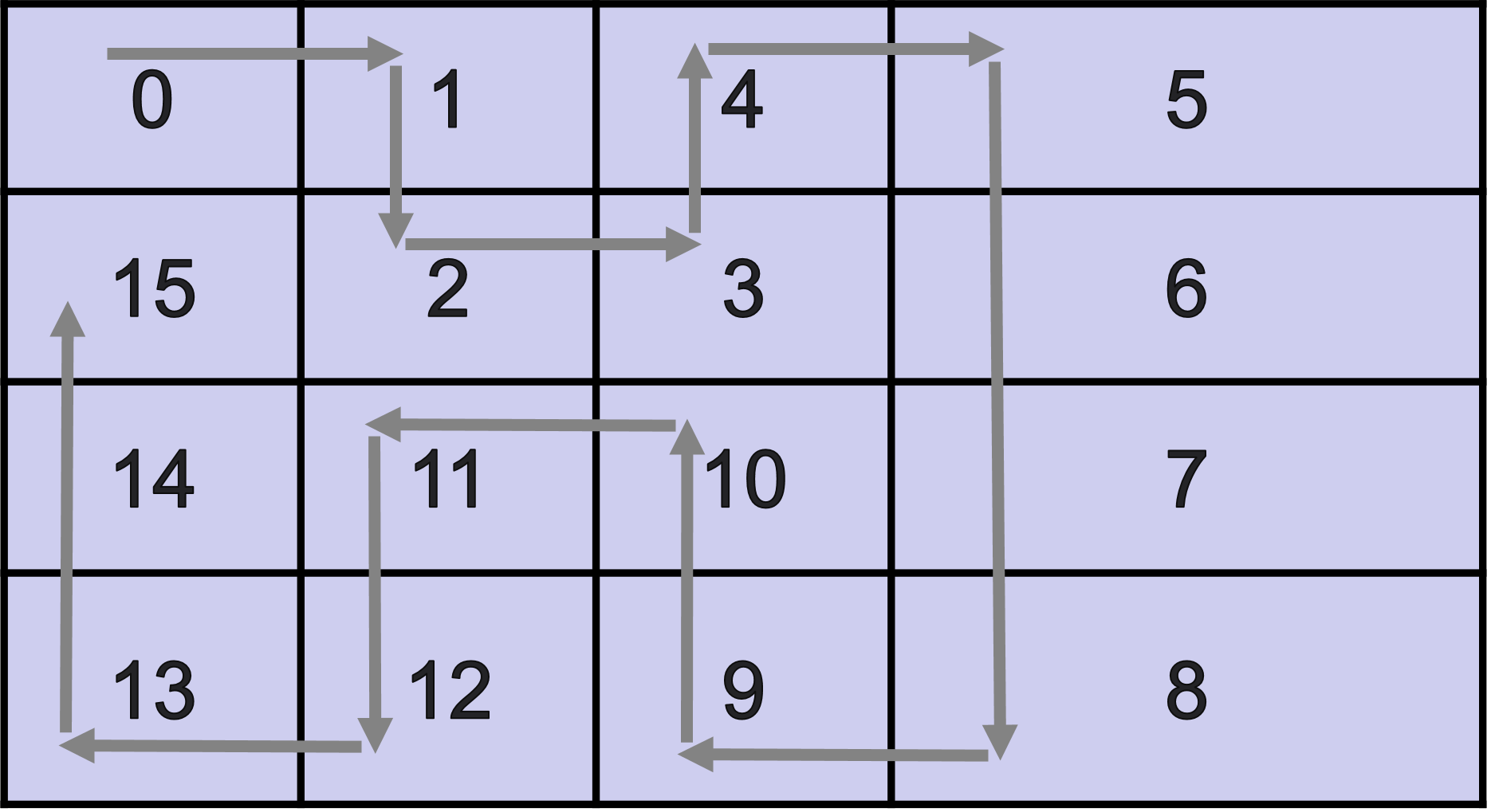}
		\caption{Hilbert-scan based coefficients merging for 1080p}\label{hilbert}
	\end{center}
\end{figure}

\subsection{Hilbert-Scan Based Coefficients Merging and Coefficients Derivation}
After corresponding region index of each pixel derived during region partition process, samples with the same region index are gathered and corresponding coefficients are calculated based on minimizing MSE principle by referring the neighboring pixels labelled in Fig.~\ref{filter shape}. Then each region has a associated Wiener filter, and after that hilbert-scan based coefficients merging is conducted according to Fig.~\ref{hilbert}. As shown in Fig.~\ref{hilbert}, a frame with the size of $1920 \times 1080$ is divided into 16 regions and the index of each region is denoted by the number labelled in it. The region with label $N$ ($N > 0$) can only merge from region $N-1$, which means that the region with label "0" can not merge from other regions, the region with label "1" can only merge from region "0" and so on. However, in same cases, region "1" may have the similar characteristics with other non-neighboring regions rather than region "2". Intuitively, regions with same labels could be merged together, but some special cases should be considered. 

\subsection{Group-based Filtering}
AVS3 defines a symmetric filter shape, which combines the $7 \times 7$ - tap cross shape and $3 \times 3$ - tap rectangular shape. The ALF filter shape supported by AVS3 is depicted in Fig.~\ref{filter shape}, where $Ct$ ($t$ being 0 to 8) represents the $t$-th filter coefficients. At encoder and decoder, each pixel sample $R(i, j)$ is filtered, resulting in pixel value $\hat{R}{(i, j)}$ as shown in Equation (1), where $K$ denotes filter length ($K = 7$), $f(k,l)$ represents filter coefficient defined in Fig.~\ref{filter shape}. 

\begin{equation}
\hat{R}{(i, j)} = \sum_{k=-K/2}^{K/2}\sum_{l=-K/2}^{K/2}f(k, l)R(i + k, j + l),
\end{equation}

In addition, to avoid over-smoothing by ALF, CTU-based on/off control is applied where a flag is signaled at CTU level to indicate whether ALF is applied to the CTU. To reduce the coefficients overhead, temporal prediction \cite{zhang2012adaptive} of previously coded filter coefficients may be used, i.e., the filter coefficients used by current frame may be from the previously coded pictures.

\section{Proposed Flexible-region Based Adaptive Loop Filter}
\subsection{Motivation}
There are mainly two problems in existing ALF technique of AVS3, one is the fixed region partition strategy, the other is the neighboring-region-only constrained merging scheme. Regarding to the former problem, on one hand, the current ALF in AVS3 tends to generate non-uniformly partitioned regions. Such unbalanced region partition mechanism may lead to low filtering performance due to unbalanced number of samples for each region. For example, each frame should be divided into 8 regions rather than 4 regions for videos with $416\times240$ resolution. For 720p, the coding performance may be better when the number of CTUs in each col set to be 3, 3, 2, 2 and the number of CTUs in each row set to be 2, 2, 1 and 1, respectively. On the other hand, fixed 16-region partition template may be not flexible enough to satisfy the complicated texture in high resolution videos, such as 4K and 8K videos.

\begin{figure}[ht]
	\begin{center}
		\noindent
		\includegraphics[width=3.4in]{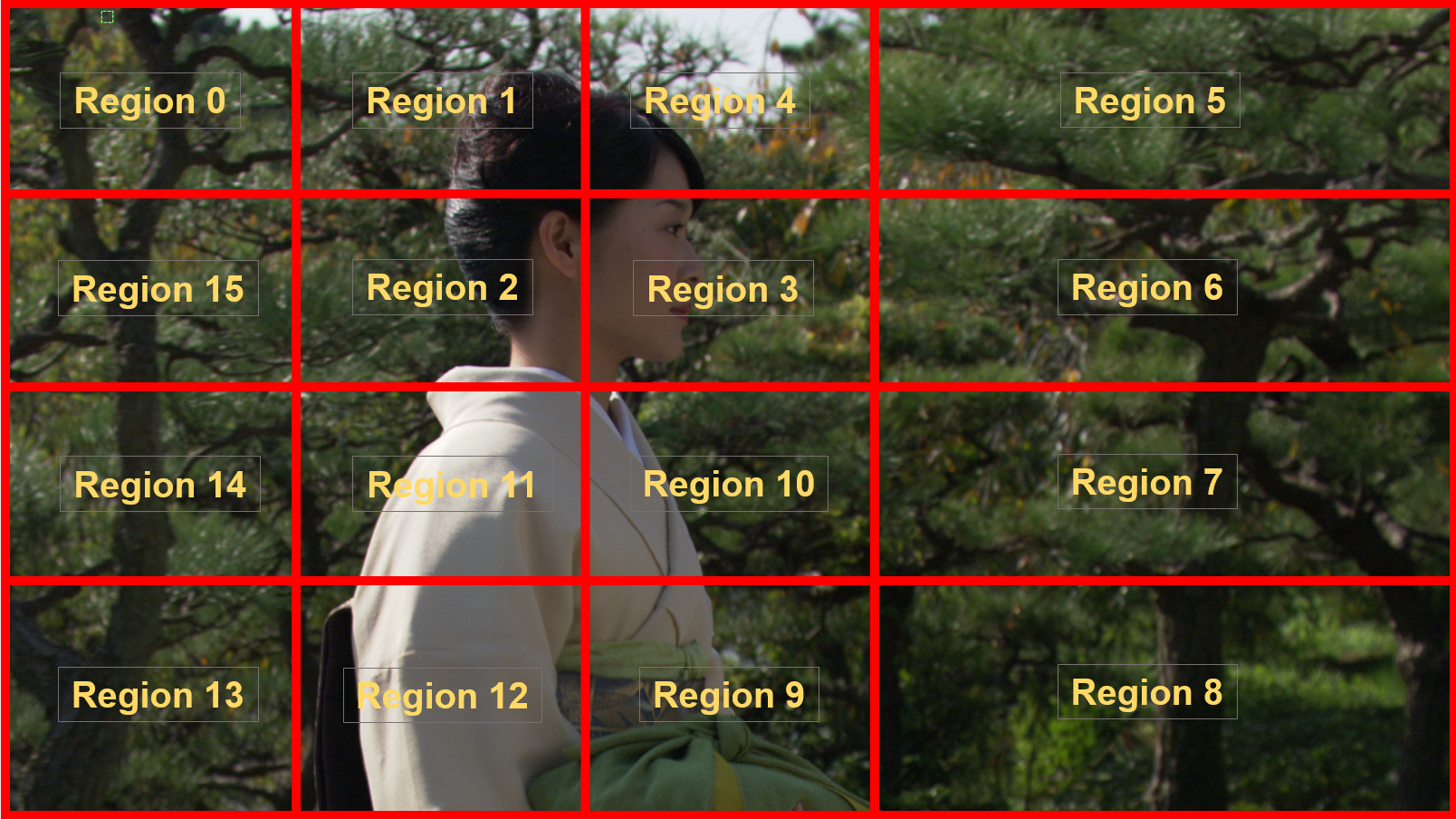}
		\caption{Region partition and coefficients merging example of ALF in AVS3}\label{partition}
	\end{center}
\end{figure}

As for the latter problem, region $N$ may have the similar characteristics with other regions rather than its spatially neighboring regions. As shown in Fig.~\ref{partition}, Region 0 is more similar with region 15/14/13, however it can only be merge with region1. In order to address these problems, we propose a flexible-region based ALF with optimized filter coefficients merging scheme to further improve the adaptability of existing ALF.

\subsection{Flexible-region partition strategy}
\begin{figure}[ht]
	\begin{center}
		\noindent
		\subfigure[$2 \times 4$]{
			\includegraphics[width=1.6in]{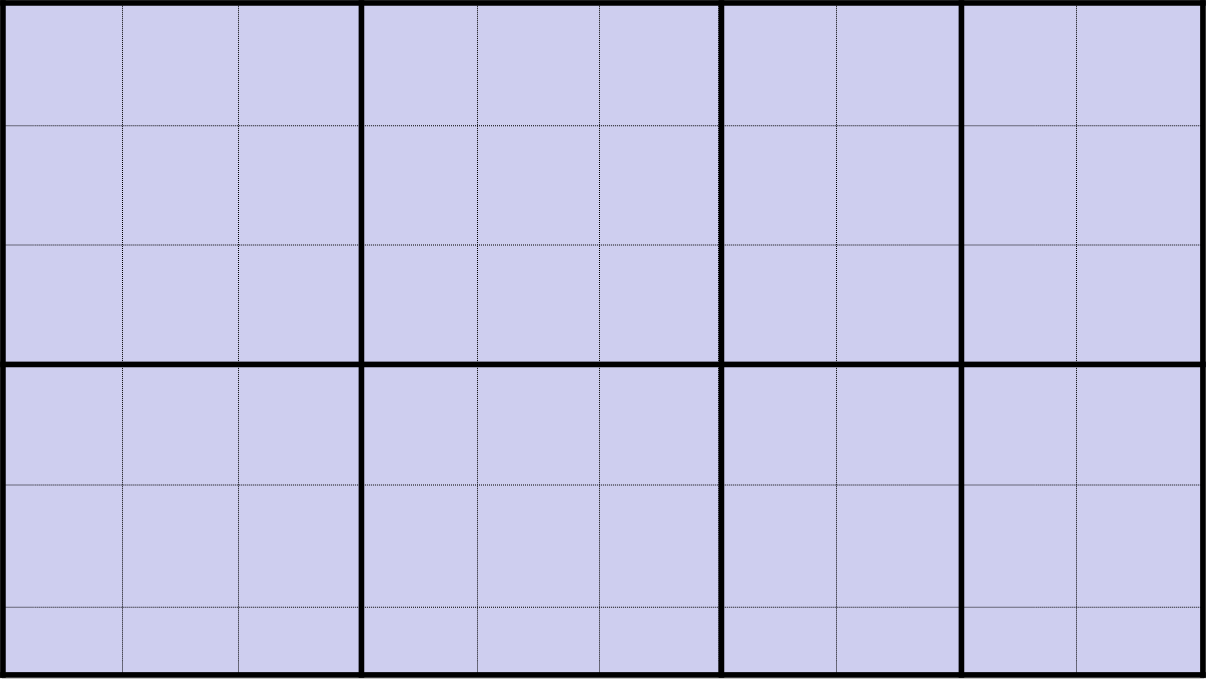}
		}
		\subfigure[$4 \times 4$]{
			\includegraphics[width=1.6in]{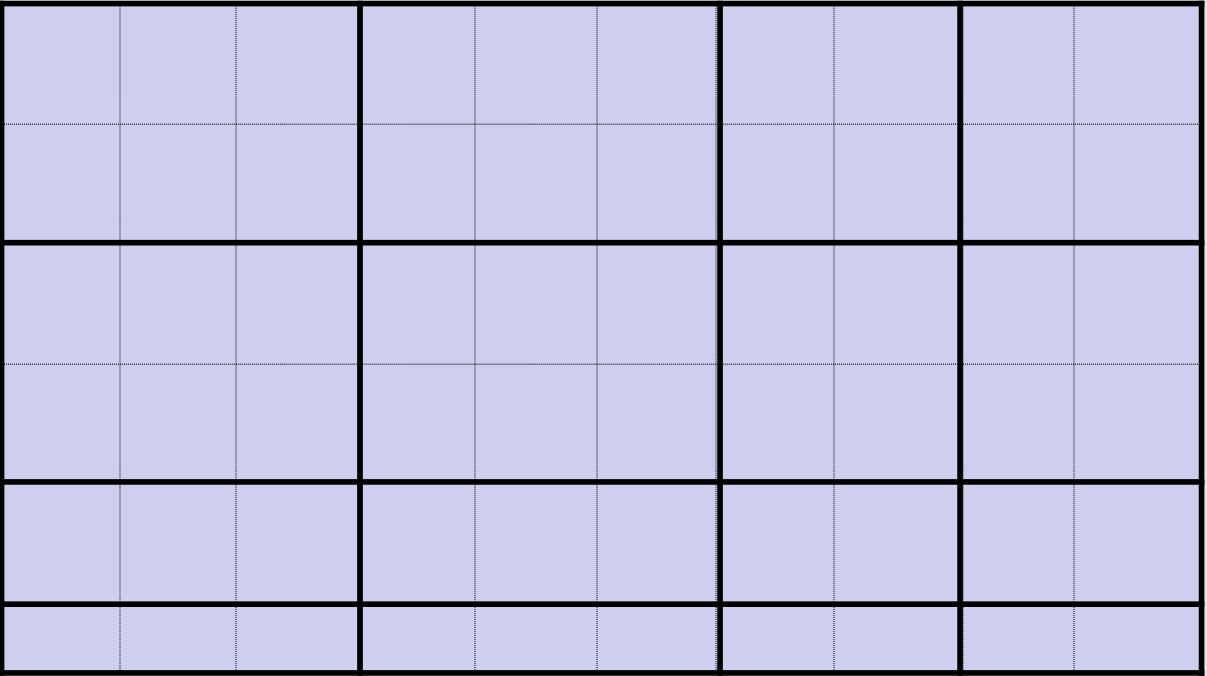}
		}
	    \subfigure[$4 \times 8$]{
		    \includegraphics[width=1.6in]{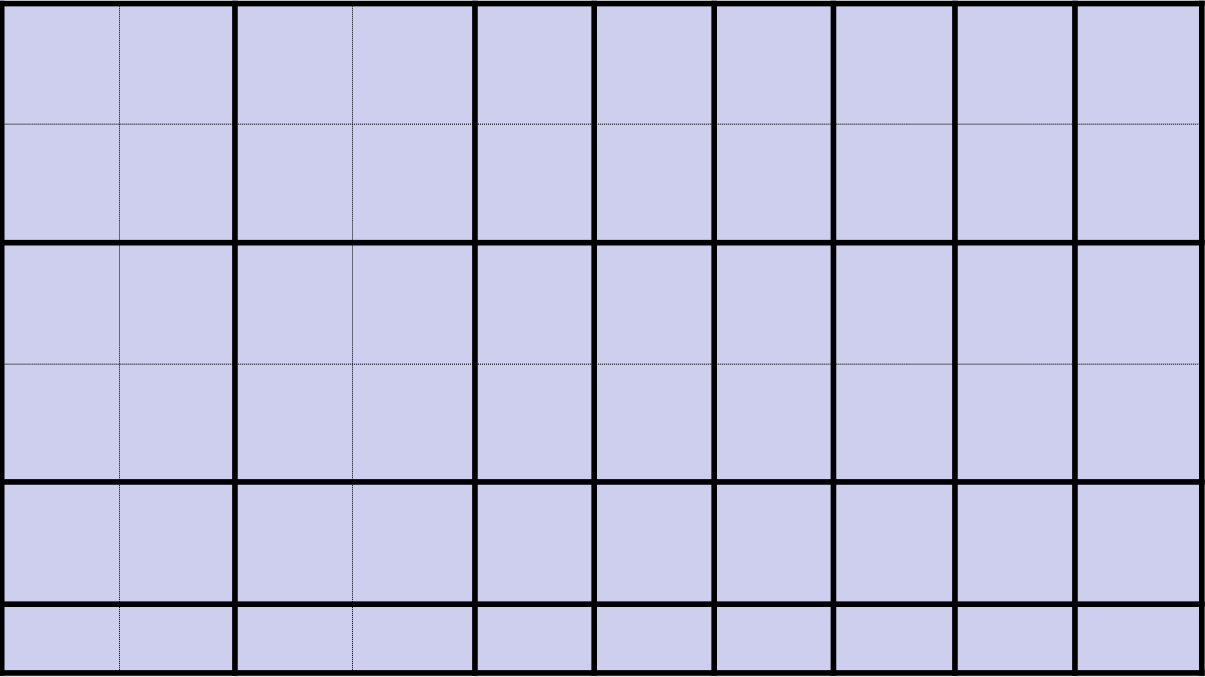}
	    }
        \subfigure[$8 \times 8$]{
	        \includegraphics[width=1.6in]{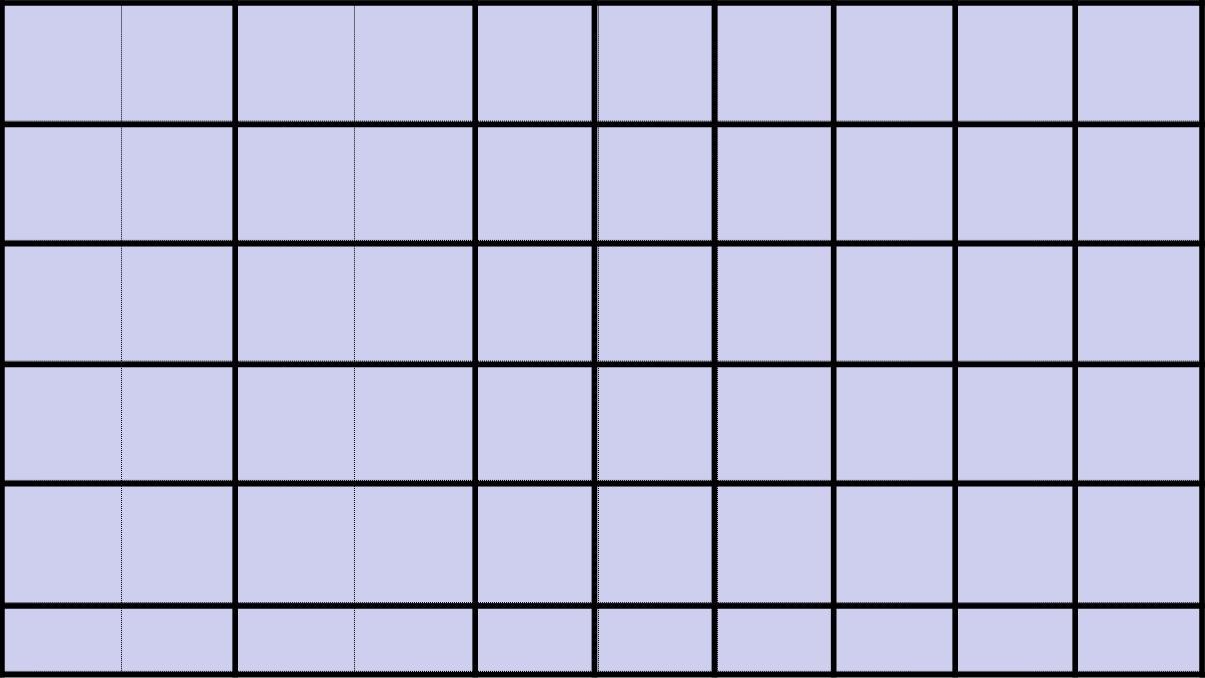}
        }
		\caption{Proposed partition strategy example for 720P}\label{partition proposed}
	\end{center}
\end{figure}

To improve the adaptability of ALF, we provide 4 types of region partition templates at encoder side, i.e., $2 \times 4$, $4 \times 4$, $4 \times 8$ and $8 \times 8$, as shown in Fig.~\ref{partition proposed}. Each frame can be divided by such four templates in encoder, then choose the best partition template in the sense of rate-distortion performance and signal its corresponding index to decoder. By using the proposed method, each region can be divided as uniformly as possible. 

\subsection{Non-local based filter coefficients merging scheme}
In order to conduct filter merging process more efficiently, we enable the ALF filters merging between any regions in each frame. However, exhaustively performing merge decision between all available regions is unrealistic because of the high complexity caused. Hence, we apply fast distortion estimation algorithm to estimate the most reliable merging mode. For each frame with $R$ regions, there are $R$ ALF filters at most. In order to derive the best merging result and the best filter coefficients, $R$ times iteration should be conducted. For $i_{th}$ $(0 < i\le R)$ iteration, merge the $R$ regions into $i$. Therefore, there is no increase in the number of iteration. 

\begin{figure}[ht]
	\begin{center}
		\noindent
		\subfigure[$QP=27$]{
			\includegraphics[width=1.6in]{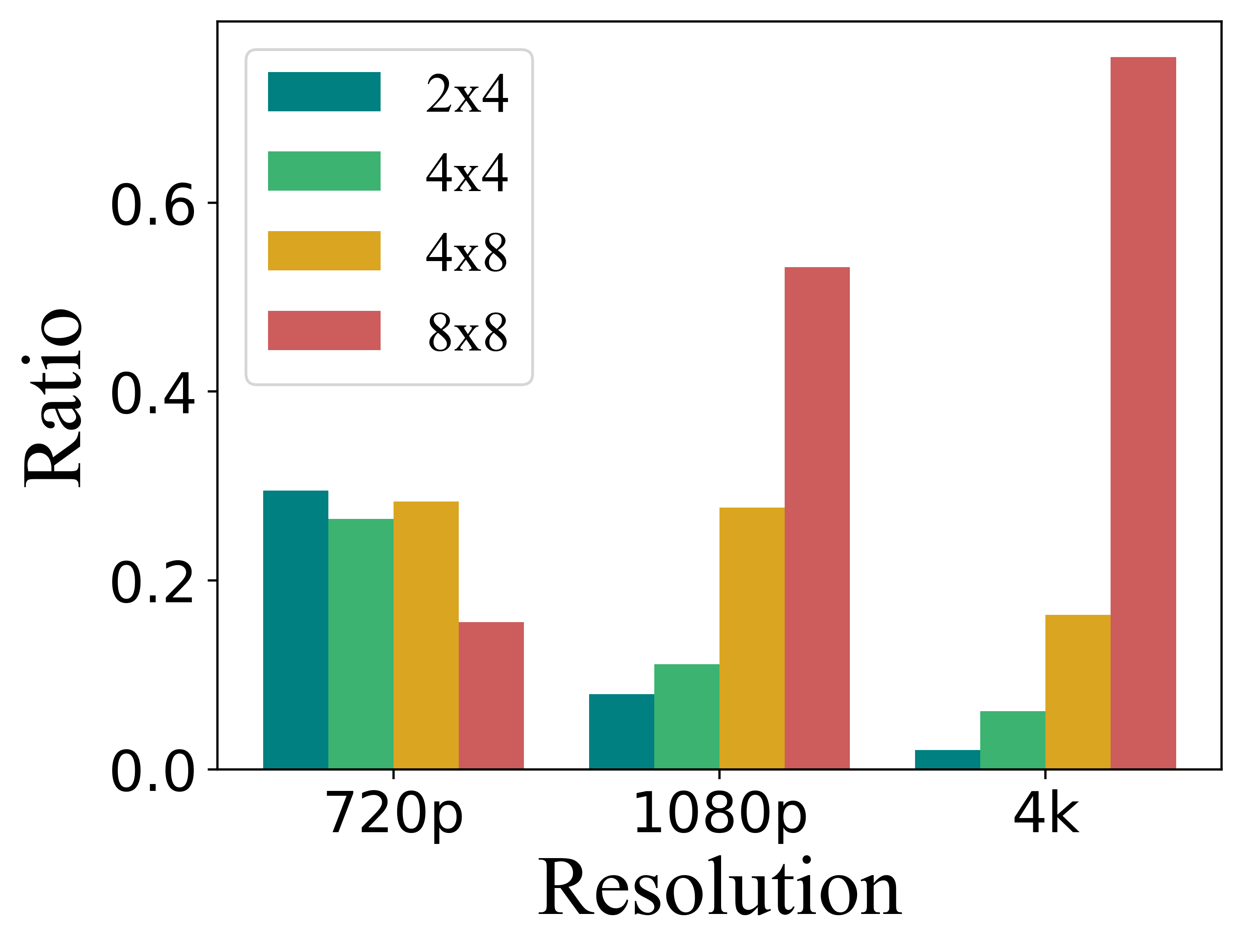}
		}
		\subfigure[$QP=32$]{
			\includegraphics[width=1.6in]{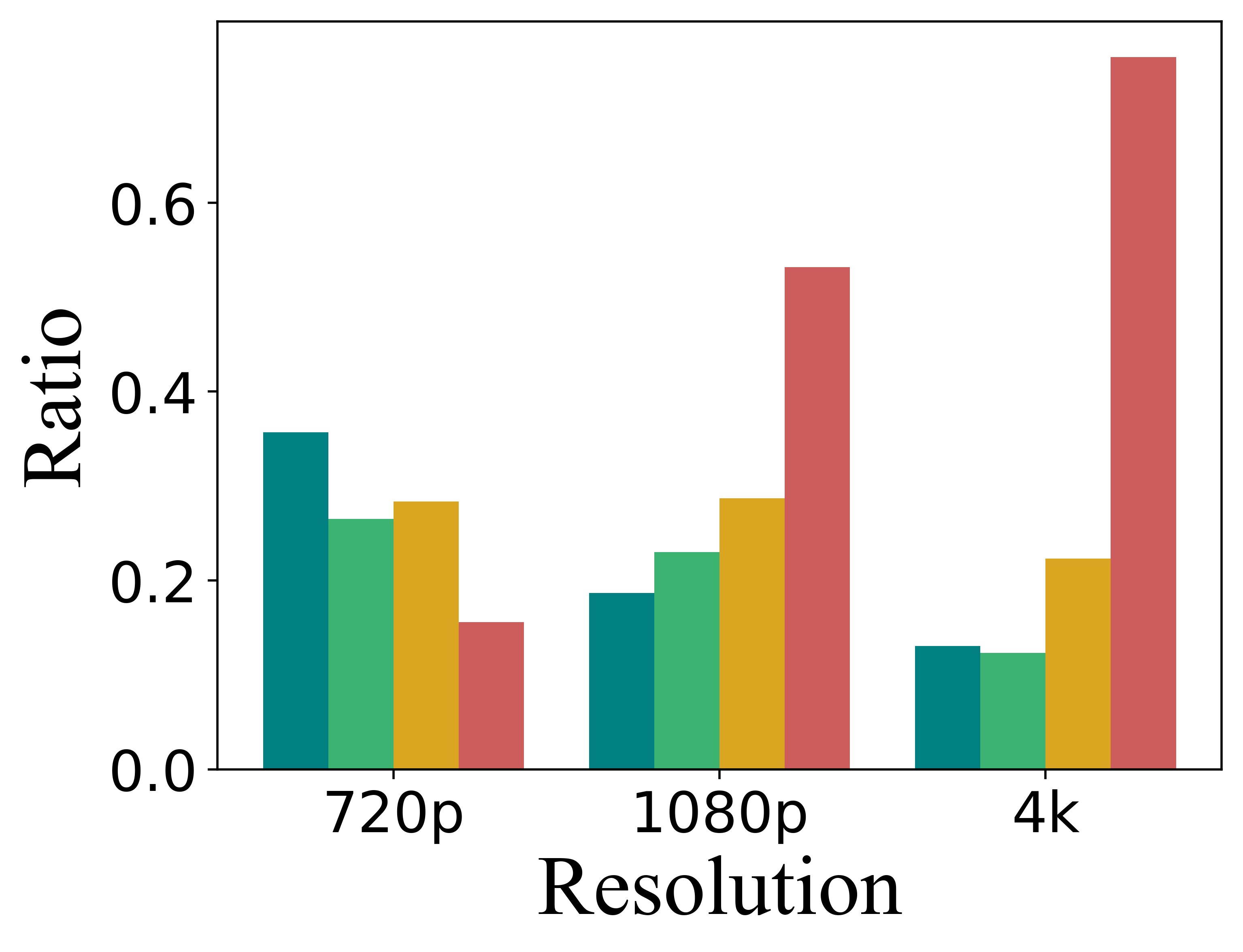}
		}
		\subfigure[$QP=38$]{
			\includegraphics[width=1.6in]{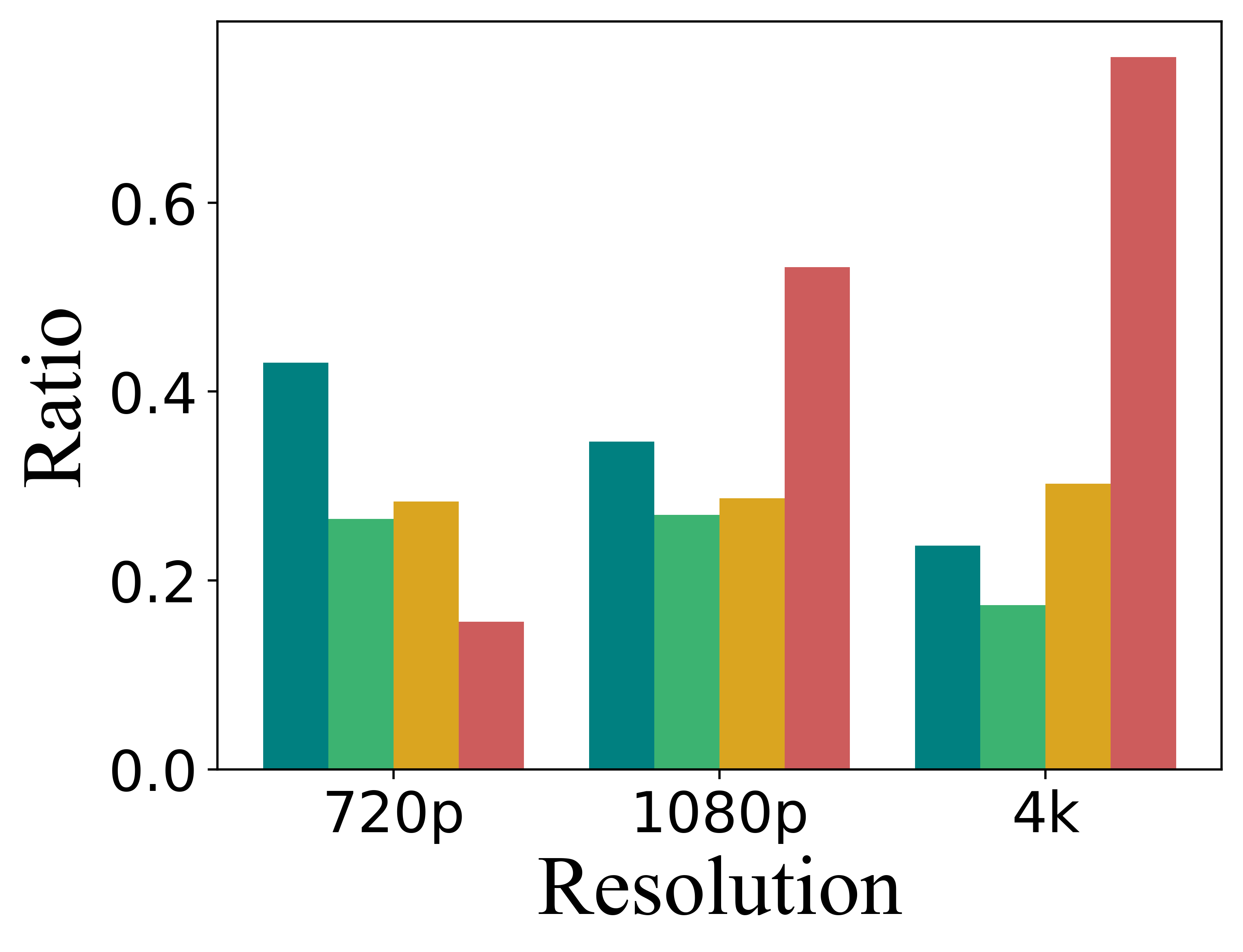}
		}
		\subfigure[$QP=45$]{
			\includegraphics[width=1.6in]{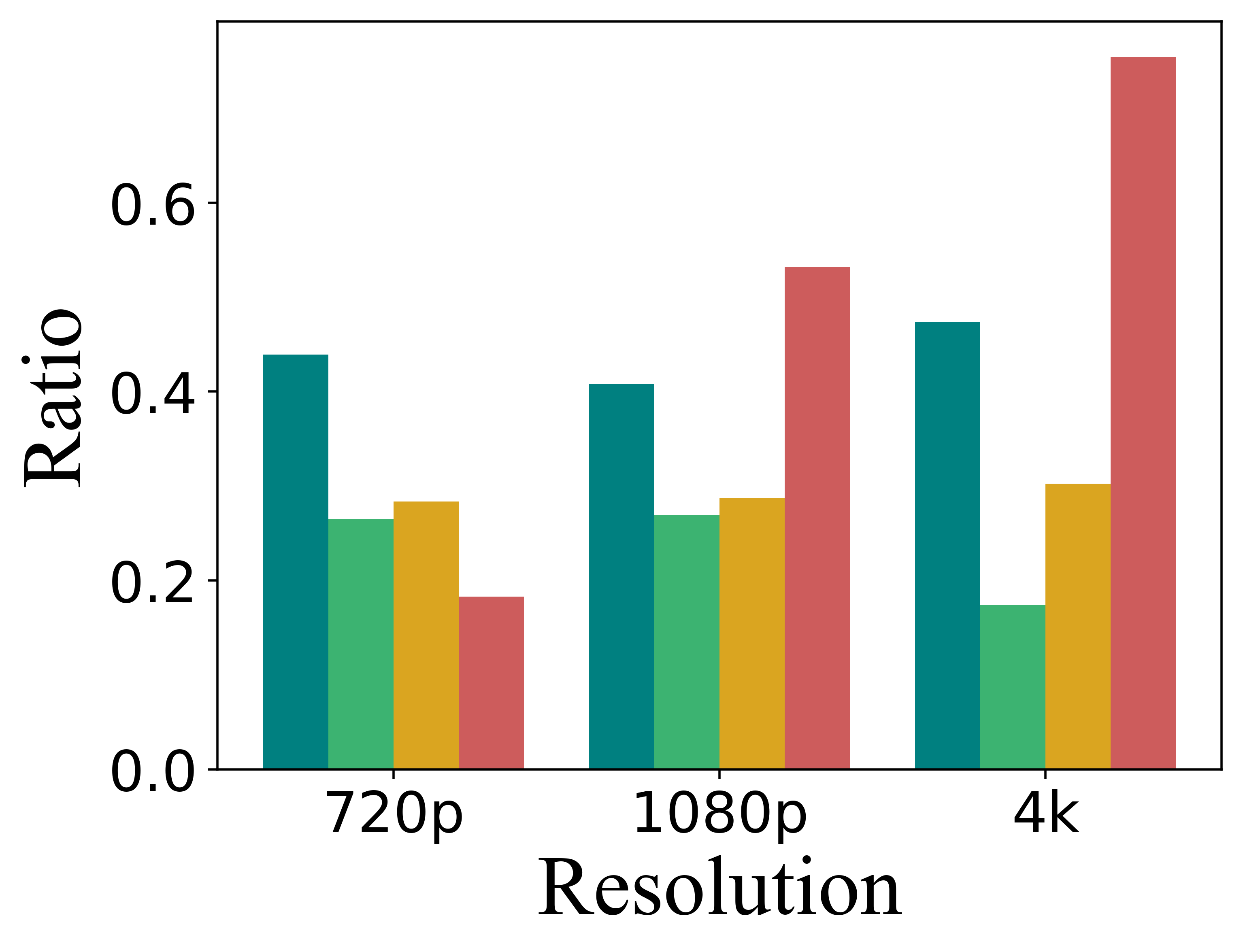}
		}
		\caption{Selection probability of different partition templates under RA configuration}\label{partition distribute}
	\end{center}
\end{figure}

According to the statistical results generated by all the test sequences in Table \ref{Performance} under AVS3 common test conditions \cite{CFP} with RA configuration shown in Fig.~\ref{partition distribute}, we observed that high resolution videos tend to select the partition template with more partition regions. While, low resolution videos are more likely to use a partition template with less regions. In addition, for high QPs, as the image quality before ALF is better than low QPs and texture details are more complicated, partition modes such as $4 \times 8$ and $8 \times 8$ are more likely to be selected. 

\makeatletter
\newcommand{\thickhline}{%
    \noalign {\ifnum 0=`}\fi \hrule height 1pt
    \futurelet \reserved@a \@xhline
}

\begin{table*}[htbp!]
	\centering
	\begin{center}
		\caption{Experimental results of Proposed Flexible-Region based ALF, Anchor: HPM-3.3 w/o ALF} \label{Performance}
		\begin{tabular}{m{1.8cm}<{\centering}|m{2.2cm}<{\centering}|m{1.7cm}<{\centering}|m{1.7cm}<{\centering}|m{1.7cm}<{\centering}|m{1.7cm}<{\centering}|m{1.7cm}<{\centering}|m{1.7cm}<{\centering}}
			%\toprule[1.5pt]
			\thickhline
			\hline
			\multirow{2}{*}{\textbf{Resolution}}&
            \multirow{2}{*}{\textbf{Sequence}}&
			\multicolumn{3}{c|}{\rule{0pt}{8pt} \textbf{FRALF compared to HPM-3.3 with ALF ON}} & \multicolumn{3}{c}{\textbf{FRALF compared to HPM-3.3 with ALF OFF}} \\
			\cline{3-8} & & 
			\textbf{AI}& \textbf{RA}& \textbf{LDB}& \textbf{AI}& \textbf{RA}& \textbf{LDB}\\
			
			\hline
			\multirow{3}{*}{3840$\times$2160}
			& \rule{0pt}{8pt}\textit{Tango2}           & -0.00\% &  0.34\% & -0.09\%  & -2.67\% & -2.64\%  & -3.51\% \\
			%& \rule{0pt}{8pt} \textit{ParkRunning3}    & -0.39\% & -xxxx\% & -0.38\%  & -6.22\% & -xxxx\%  & -7.31\% %\\
			& \rule{0pt}{8pt} \textit{Campfire}        & -0.10\% & -0.12\% & -0.29\%  & -0.92\% & -2.46\%  & -2.25\% \\
			& \rule{0pt}{8pt} \textit{DaylightRoad2}   & -0.65\% & -0.70\% & -0.84\%  & -2.85\% & -4.36\%  & -3.96\%  \\
			
			\hline
			\multirow{4}{*}{1920$\times$1080}
			& \rule{0pt}{8pt} \textit{Cactus}          & -0.40\% & -0.48\% & -0.27\%  & -2.81\% & -3.61\%  & -2.15\%  \\
			& \rule{0pt}{8pt} \textit{BasketballDrive} & -0.55\% & -0.61\% & -0.58\%  & -1.65\% & -2.81\%  & -2.82\%  \\
			& \rule{0pt}{8pt} \textit{MarketPlace}     & -0.29\% & -0.37\% & -0.09\%  & -3.70\% & -5.78\%  & -5.04\%  \\
			& \rule{0pt}{8pt} \textit{RitualDance}     & -0.31\% & 0.02\%  &  0.09\%  & -3.97\% & -2.23\%  & -2.06\%   \\
			
			\hline
			\multirow{4}{*}{1280$\times$720}
			& \rule{0pt}{8pt} \textit{City}            & -0.37\% & -0.71\% & -1.07\%  & -3.82\% & -8.61\%  & -7.70\%  \\
			& \rule{0pt}{8pt} \textit{Crew}            & -0.08\% & -0.16\% & -0.12\%  & -2.73\% & -3.31\%  & -3.43\%  \\
			& \rule{0pt}{8pt} \textit{vidyo1}          & -0.32\% & -0.36\% & -0.22\%  & -2.78\% & -2.43\%  & -2.43\%  \\
			& \rule{0pt}{8pt} \textit{vidyo3}          & -0.35\% & -0.77\% & -0.52\%  & -3.72\% & -6.27\%  & -5.76\%   \\
			\hline
			
			\multicolumn{2}{c|}{\rule{0pt}{8pt} Average}   & -0.31\% & -0.36\% & -0.36\% & -2.87\% & -4.05\% & -3.74\% \\
			\hline
			\multicolumn{2}{c|}{\rule{0pt}{8pt} Enc Time Ratio (\%)} & 101.06\% & 101.23\% & 101.44\% & 102.78\% & 101.99\% & 101.81\% \\
			\hline
			\multicolumn{2}{c|}{\rule{0pt}{8pt} Dec Time Ratio (\%)} & 100.56\% & 100.12\% & 100.91\% & 103.55\% & 102.92\% & 102.03\% \\
			\thickhline
		\end{tabular}
	\end{center}
\end{table*}

\section{Experimental Results}
In this section, we test the performance of the proposed flexible-region based ALF by implementing it into HPM-3.3. The experiments were performed under AVS3 common test condition\cite{CFP}. Three configurations used in AVS3 are tested: AI, RA and LDB. Coding performance is measured by Bjontegaard's method \cite{BDrate} in terms of BD-rate (Y component). We also evaluate the computational complexity by comparing the running time of encoder and decoder respectively. The executable files are compiled by Microsoft Visual Studio 2017, 64bit. The encoder tests run on Linux system with 64 bit and CPU in the test is Intel Xeon E5-2697A V4 @ 2.60GHz. While, the decoder tests run on Windows system with 64 bit and CPU in the test is Intel(R) Core™ i7-8700 @ 3.20GHz.

\begin{figure}[ht]
	\begin{center}
		\noindent
		\includegraphics[width=3.4in]{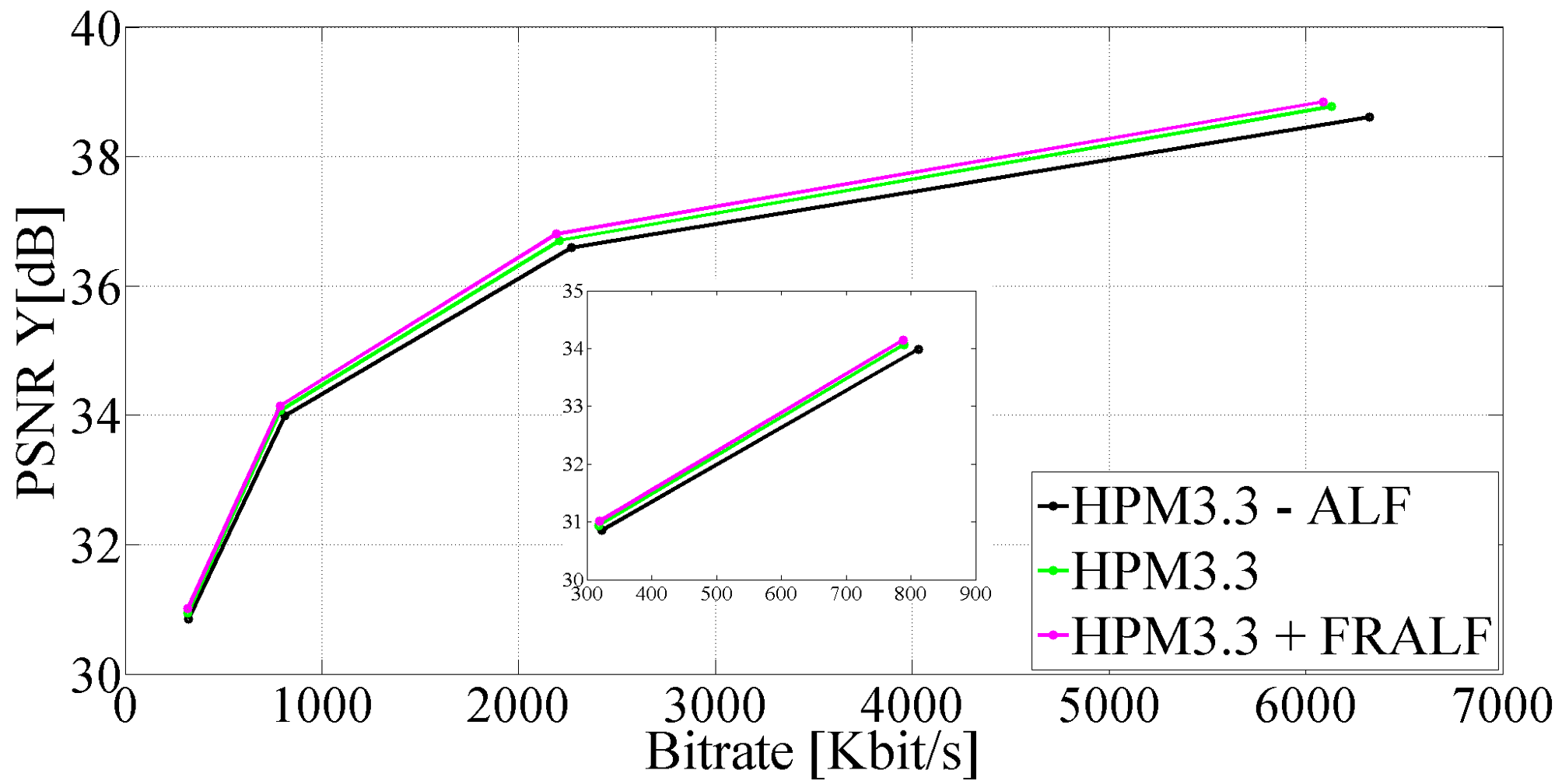}
		\caption{Rate-distortion curves of  \emph{vidyo1} LDB configuration. }\label{BD-rate}
	\end{center}
\end{figure}

\begin{figure}[ht]
	\begin{center}
		\noindent
		\subfigure[HPM3.3 - ALF OFF]{
			\includegraphics[width=1.0in]{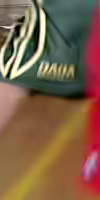}
		}
		\subfigure[HPM3.3]{
			\includegraphics[width=1.0in]{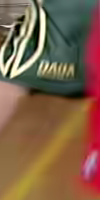}
		}
		\subfigure[HPM3.3 + FRALF]{
			\includegraphics[width=1.0in]{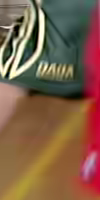}
		}
		\caption{Visual comparisons between Anchor and proposed FRALF of frames from \emph{BasketballDrive}, (a) HPM3.3(LDB), ALF OFF, QP = 45, SSIM = 0.972 (b) HPM3.3(LDB), ALF ON, QP = 45, SSIM = 0.972 (c) HPM3.3(LDB), FRALF, QP = 45, SSIM = 0.973 }\label{subjective}
	\end{center}
\end{figure}

TABLE \ref{Performance} summarizes the results of proposed method. There are two tests conducted, including FRALF compared to HPM-3.3 and FRALF compared to HPM-3.3 with ALF disabled. In test one, anchor is generated by HPM-3.3 with default configuration. In test two, anchor is HPM-3.3 with ALF disabled. It can be seen that the proposed method can achieve 0.31\%, 0.36\% and 0.36\% bit-rate reduction when ALF on and 2.87\%, 4.05\% and 3.74\% bit-rate saving w/o ALF with minor encoding /decoding time increase in AI, RA and LDB configurations, respectively. Note that the bit rate reduction for RA and LDB configurations is higher. The reason is that
the quality of reference pictures is improved by FRALF, which is beneficial to efficiently compress the following frames. FRALF could bring up to 1.07\% coding gain
on top of the existing ALF for test sequence \emph{City} under LDB
configuration and the overall bit rate reduction of adaptive loop
filter is 7.70\% in this case. The
rate-distortion curves of this sequence are depicted in Fig.~\ref{BD-rate}, where ‘HPM3.3-ALF’ represents the results achieved by HPM-3.3 with ALF disabled, ‘HPM3.3’ represents the results of HPM-3.3 with existing ALF enabled and ‘HPM3.3+FRALF’ denotes the proposed method. From this figure, we can see that FRALF is able to improve coding efficiency for both low and high qualities of test sequence. In addition to objective quality improvement, FRALF can also improve subjective performance as show in Fig.~\ref{subjective}. Compared to HPM3.3 with ALF disabled, FRALF can improve subjective quality obviously. 

\section{Conclusion}
In this paper, a flexible region partition based adaptive loop filter technique is proposed to improve the coding efficiency of ALF. In the proposed method, multiple region partition templates and non-local based filtering merge scheme are introduced to satisfy the increasing requirements of high resolution and multi-scene video coding, in which the local texture characteristics and non-local similarities are fully considered synthetically. Experimental results demonstrate that our proposed algorithm can achieve 0.31\%, 0.36\% and 0.36\% bit-rate reduction on average for AI, RA and LDB configurations, respectively, which improves the adaptability of ALF with marginal complexity increase.

\bibliography{reference}

@inproceedings{r0322,
  title={{CCALF} virtual boundary issue for 4:4:4 and 4:2:2 format},
  author={Meng, Xuewei and Zheng, Xiaozhen and Wang, Shanshe and Ma, Siwei},
  booktitle={Joint Video Experts Team (JVET), document JVET-R0322},
  year={2020}
}

@inproceedings{Parallel,
	title={{AHG} 10: {ALF} and {CCALF} encoder parallel design},
	author={Meng, Xuewei and others},
	booktitle={Joint Video Experts Team (JVET), document JVET-R0328},
	year={2020}
}

@inproceedings{NonlinearALF-2,
  title={Non-{CE5}: Modification of clipping value signalling for {ALF}},
  author={Meng, Xuewei and others},
  booktitle={Joint Video Experts Team (JVET) of ITU-T SG, JVET-O0430, 15th JVET Meeting},
  year={2019}
}

@inproceedings{R0225,
	title={{AHG}9: On {ALF/CC-ALF} high level syntax},
	author={Meng, Xuewei and Zheng, Xiaozhen and Wang, Shanshe and Ma, Siwei},
	booktitle={JVET document, JVET-R0225},
	year={2020}
}

@inproceedings{nonlocal2,
  title={Description of {SDR} video coding technology proposal by {DJI} and {P}eking {U}niversity},
  author={Wang, Zhao and Meng, Xuewei and Jia, Chuanmin and Cui, Jing and Wang, Suhong and Wang, Shanshe and Ma, Siwei},
  booktitle={Document of Joint Video Experts Team, JVET-J0011, 10th JVET Meeting},
  year={2018}
}

@inproceedings{nonlocal3,
  title={Non-local Structure-based Filter with integer operation},
  author={Meng, Xuewei and Jia, Chuanmin and Wang, Zhao and Ma, Siwei and Zheng, Xiaozhen},
  booktitle={Document of Joint Video Experts Team, JVET-J0071, 10th JVET Meeting},
  year={2018}
}

@inproceedings{nonlocal4,
  title={{CE}2: Non-local structure-based filter},
  author={Meng, Xuewei and others},
  booktitle={Document of Joint Video Experts Team, JVET-K0160, 11th JVET Meeting},
  year={2018}
}

@inproceedings{OnePass,
	title={{AHG} 10: {O}ne-pass {CCALF}},
	author={Meng, Xuewei and Zheng, Xiaozhen and Wang, Shanshe and Ma, Siwei},
	booktitle={Document of Joint Video Experts Team, JVET-R0327, 18th JVET Meeting},
	year={2020}
}

@article{CC-ALF,
	title={{CC-ALF}: {I}ntegrated {T}ext for the {C}ross {C}omponent {A}daptive {L}oop {F}ilter},
	author={Misra, Kiran and Bossen, Frank and Segall, Andrew and others},
	journal={JVET-Q0795, 17th JVET Meeting, Brussels, BE},
	year={2020}
}

@inproceedings{CCALF1,
  title={{CE}5-related: On {CC-ALF} slice and picture header syntax},
  author={Meng, Xuewei and Zheng, Xiaozhen and Wang, Shanshe and Ma, Siwei},
  booktitle={Document of Joint Video Experts Team, JVET-Q0326, 17th JVET Meeting},
  year={2019}
}

@inproceedings{CCALF2,
  title={{CE}5-related: High level syntax modifications for {CCALF} (combination of {JVET-Q0253} and {JVET-Q0520})},
  author={A. M. Kotra and S. Esenlik and B. Wang and H. Gao and others},
  booktitle={Document of Joint Video Experts Team, JVET-Q0782, 17th JVET Meeting},
  year={2019}
}

@inproceedings{NonlinearALF-1,
  title={Non-{CE5}: Unification of non-linear {ALF} luma/chroma clipping parameters},
  author={Meng, Xuewei and Zheng, Xiaozhen and Wang, Shanshe and Ma, Siwei},
  booktitle={Document of Joint Video Experts Team, JVET-O0437, 15th JVET Meeting},
  year={2019}
}

@misc{meng2026optimizedadaptiveloopfilter,
      title={Optimized Adaptive Loop Filter in Versatile Video Coding}, 
      author={Xuewei Meng and Jiaqi Zhang and Chuanmin Jia and Xinfeng Zhang and Shanshe Wang and Siwei Ma},
      year={2026},
      eprint={2607.05737},
      archivePrefix={arXiv},
      primaryClass={cs.CV},
      url={https://arxiv.org/abs/2607.05737}, 
}

@inproceedings{meng2021optimized,
  title={Optimized adaptive loop filter in versatile video coding},
  author={Meng, Xuewei and Zhang, Jiaqi and Jia, Chuanmin and Xinfeng, Zhang and Shanshe, Wang and Siwei, Ma},
  booktitle={Data Compression Conference (DCC)},
  pages={359--359},
  year={2021},
  organization={IEEE}
}

@incollection{AVSOverview,
	title={An overview of {AVS}2 standard},
	author={Gao, Wen and Ma, Siwei},
	booktitle={Advanced Video Coding Systems},
	pages={35--49},
	year={2014},
	publisher={Springer}
}

@article{AVS2?,
	title={{AVS2}? Making video coding smarter [standards in a Nutshell]},
	author={Ma, Siwei and Huang, Tiejun and Reader, Cliff and Gao, Wen},
	journal={IEEE Signal Processing Magazine},
	volume={32},
	number={2},
	pages={172--183},
	year={2015},
	publisher={IEEE}
}

@inproceedings{AVS2framework,
	title={Framework of {AVS}2-video coding},
	author={He, Zhichu and Yu, Lu and Zheng, Xiaozhen and Ma, Siwei and He, Yun},
	booktitle={IEEE International Conference on Image Processing},
	pages={1515--1519},
	year={2013},
	organization={IEEE}
}

@inproceedings{meng2018optimized,
	title={Optimized Non-local In-Loop Filter for Video Coding},
	author={Meng, Xuewei and Jia, Chuanmin and Wang, Shanshe and Zheng, Xiaozhen and Ma, Siwei},
	booktitle={Picture Coding Symposium (PCS)},
	pages={233--237},
	year={2018},
	organization={IEEE}
}

@inproceedings{GALF,
	title={Geometry transformation-based adaptive in-loop filter},
	author={Karczewicz, Marta and Zhang, Li and Chien, Wei-Jung and Li, Xiang},
	booktitle={Picture Coding Symposium (PCS)},
	pages={1--5},
	year={2016},
	organization={IEEE}
}

@article{HEVC,
    title={Overview of the {H}igh {E}fficiency {V}ideo {C}oding ({HEVC}) standard},
    author={Sullivan, Gary J and Ohm, Jens-Rainer and Han, Woo-Jin and Wiegand, Thomas},
    journal={IEEE Transactions on Circuits and Systems for Video Technology},
    volume={22},
    number={12},
    pages={1649--1668},
    year={2012},
    publisher={IEEE}
}

@article{DF,
	title={Adaptive deblocking filter},
	author={List, Peter and Joch, Anthony and Lainema, Jani and Bjontegaard, Gisle and Karczewicz, Marta},
	journal={IEEE Transactions on Circuits and Systems for Video Technology},
	volume={13},
	number={7},
	pages={614--619},
	year={2003},
	publisher={IEEE}
}

@article{SAO,
	title={Sample {A}daptive {O}ffset in the {HEVC} standard},
	author={Fu, Chih-Ming and Alshina, Elena and Alshin, Alexander and Huang, Yu-Wen and Chen, Ching-Yeh and Tsai, Chia-Yang and Hsu, Chih-Wei and Lei, Shaw-Min and Park, Jeong-Hoon and Han, Woo-Jin},
	journal={IEEE Transactions on Circuits and Systems for Video Technology},
	volume={22},
	number={12},
	pages={1755--1764},
	year={2012},
	publisher={IEEE}
}

@article{BF,
	title={Bilateral filter strength based on prediction mode},
	author={Str{\"o}m, Jacob and Wennersten, Per and Andersson, Kenneth and Enhorn, Jack},
	journal={JVET of ITU-T SG 16 WP 3 and ISO/IEC JTC1/SC29/WG11, Doc. JVET-E0032, 5th Meeting, Geneva, CH},
	year={2017}
}

@article{ALF,
	title={Adaptive loop filtering for video coding},
	author={Tsai, Chia-Yang and Chen, Ching-Yeh and Yamakage, Tomoo and Chong, In Suk and Huang, Yu-Wen and Fu, Chih-Ming and Itoh, Takayuki and Watanabe, Takashi and Chujoh, Takeshi and Karczewicz, Marta and others},
	journal={IEEE Journal of Selected Topics in Signal Processing},
	volume={7},
	number={6},
	pages={934--945},
	year={2013},
	publisher={IEEE}
}

@article{NALF,
	title={Low-Rank Based Nonlocal Adaptive Loop Filter for {H}igh {E}fficiency {V}ideo {C}ompression},
	author={Zhang, Xinfeng and Xiong, Ruiqin and Lin, Weisi and Zhang, Jian and Wang, Shiqi and Ma, Siwei and Gao, Wen},
	journal={IEEE Transactions on Circuits and Systems for Video Technology},
	volume={27},
	number={10},
	pages={2177--2188},
	year={2017},
	publisher={IEEE}
}

@inproceedings{zhang2015nonlocal,
	title={Nonlocal adaptive in-loop filter via content-dependent soft-thresholding for {HEVC}},
	author={Zhang, Xinfeng and Lin, Weisi and Wang, Shiqi and Ma, Siwei},
	booktitle={IEEE International Symposium on Multimedia (ISM)},
	pages={465--470},
	year={2015},
	organization={IEEE}
}

@article{ma2016nonlocal,
	title={Nonlocal in-loop filter: The way toward next-generation video coding?},
	author={Ma, Siwei and Zhang, Xinfeng and Zhang, Jian and Jia, Chuanmin and Wang, Shiqi and Gao, Wen},
	journal={IEEE MultiMedia},
	volume={23},
	number={2},
	pages={16--26},
	year={2016},
	publisher={IEEE}
}

@inproceedings{SANF,
	title={Structure-driven adaptive non-local filter for {H}igh {E}fficiency {V}ideo {C}oding ({HEVC})},
	author={Zhang, Jian and Jia, Chuanmin and Zhang, Nan and Ma, Siwei and Gao, Wen},
	booktitle={Data Compression Conference (DCC)},
	pages={91--100},
	year={2016},
	organization={IEEE}
}

@inproceedings{zhang2012adaptive,
	title={Adaptive loop filter with temporal prediction},
	author={Zhang, Xinfeng and Xiong, Ruiqin and Ma, Siwei and Gao, Wen},
	booktitle={Picture Coding Symposium},
	pages={437--440},
	year={2012},
	organization={IEEE}
}

@article{jia2019content,
	title={Content-Aware Convolutional Neural Network for In-loop Filtering in {H}igh {E}fficiency {V}ideo {C}oding},
	author={Jia, Chuanmin and Wang, Shiqi and Zhang, Xinfeng and Wang, Shanshe and Liu, Jiaying and Pu, Shiliang and Ma, Siwei},
	journal={IEEE Transactions on Image Processing},
	year={2019},
	publisher={IEEE}
}

@article{DF_AVS,
	title={Improvement of de-blocking filter in {AVS}2},
	author={He, Jianqiang and Ma, Siwei},
	journal={AVS Doc. AVS-M3013},
	year={2012}
}

@article{SAO_AVS,
	title={Sample adaptive offset for {AVS}2},
	author={Chen, Jie and Lee, Sunil and Kim, Chanyul and Fu, Chih-Ming and Huan, Yu-Wen  and Lei, Shawmin},
	journal={AVS Doc. AVS-M3197},
	year={2013}
}

@article{ALF_AVS,
	title={Adaptive loop filter for {AVS}2},
	author={Zhang, Xinfeng and Si, Junjun and Wang, Shanshe and Ma, Siwei and Cai, Jiayang and Chen, Qinghua and Huang, Yu-Wen and Lei, Shawmin},
	journal={AVS Doc. AVS-M3292},
	year={2014}
}

@article{CFP,
	title={Common test conditions of {AVS3-P2}},
	author={Fan, Kui},
	journal={AVS Doc. AVS-N2654},
	year={2019}
}

@article{HPM,
	title={{HPM}: the new reference software of {AVS3}},
	author={Fan, Kui and Xie, Xi and Wang, Zhenyu and Xu, Guisen and Wang, Ronggang and Lu, Xiaomu and Chen, Huanbang and Zhao, Yin and Yang, Haitao},
	journal={AVS Doc. AVS-M4510},
	year={2018}
}

@article{BDrate,
	title={Calculation of average {PSNR} differences between {RD}-curves},
	author={Bjontegarrd, Gisle},
	journal={VCEG-M33, 13th VCEG Meeting, Austin, TX, USA},
	year={2001}
}

\end{document}